\documentclass[12pt,a4paper]{article}

\usepackage{ifthen} 
\newboolean{pdflatex}
\setboolean{pdflatex}{true} 

\newboolean{articletitles}
\setboolean{articletitles}{true} 

\newboolean{uprightparticles}
\setboolean{uprightparticles}{false} 

\usepackage{float}
\usepackage{caption}

\def\paperauthors{LHCb collaboration} 
\def\paperasciititle{Measurement of the Xib- production rate} 
\def\papertitle{Measurement of the mass and production rate of $\Xibm$ baryons} 
\def\paperkeywords{{High Energy Physics}, {LHCb}} 
\def\papercopyright{\the\year\ CERN for the benefit of the LHCb collaboration} 
\def\paperlicence{CC-BY-4.0 licence}
\def\paperlicenceurl{https://creativecommons.org/licenses/by/4.0/}

\usepackage[top=1in, bottom=1.25in, left=1in, right=1in]{geometry}

\usepackage{microtype}
\usepackage{lineno}  
\usepackage{xspace} 
\usepackage{caption} 

\usepackage{graphicx}  
\usepackage{color}
\usepackage{colortbl}
\graphicspath{{./figs/}} 
\DeclareGraphicsExtensions{.pdf,.PDF,png,.PNG}

\usepackage{amsmath} 
\usepackage{amssymb}
\usepackage{amsfonts}
\usepackage{upgreek} 

\newcommand*\patchAmsMathEnvironmentForLineno[1]{%
\expandafter\let\csname old#1\expandafter\endcsname\csname #1\endcsname
\expandafter\let\csname oldend#1\expandafter\endcsname\csname
end#1\endcsname
 \renewenvironment{#1}%
   {\linenomath\csname old#1\endcsname}%
   {\csname oldend#1\endcsname\endlinenomath}%
}
\newcommand*\patchBothAmsMathEnvironmentsForLineno[1]{%
  \patchAmsMathEnvironmentForLineno{#1}%
  \patchAmsMathEnvironmentForLineno{#1*}%
}
\AtBeginDocument{%
\patchBothAmsMathEnvironmentsForLineno{equation}%
\patchBothAmsMathEnvironmentsForLineno{align}%
\patchBothAmsMathEnvironmentsForLineno{flalign}%
\patchBothAmsMathEnvironmentsForLineno{alignat}%
\patchBothAmsMathEnvironmentsForLineno{gather}%
\patchBothAmsMathEnvironmentsForLineno{multline}%
\patchBothAmsMathEnvironmentsForLineno{eqnarray}%
}

\usepackage{hyperxmp}

\usepackage[pdftex,
            pdfauthor={\paperauthors},
            pdftitle={\paperasciititle},
            pdfkeywords={\paperkeywords},
            pdfcopyright={Copyright (C) \papercopyright},
            pdflicenseurl={\paperlicenceurl}]{hyperref}

\usepackage[all]{hypcap} 

\usepackage{xspace} 
\usepackage{upgreek}

\def\lhcb   {\mbox{LHCb}\xspace}

\def\MagUp {\mbox{\em Mag\kern -0.05em Up}\xspace}

\ifthenelse{\boolean{uprightparticles}}%
{

 \def\Pmu         {\ensuremath{\upmu}\xspace}                 
 \def\Pnu         {\ensuremath{\upnu}\xspace}                 
                  
 \def\Ppi         {\ensuremath{\uppi}\xspace}

 \def\Ppsi        {\ensuremath{\uppsi}\xspace}

 \def\PDelta      {\ensuremath{\Delta}\xspace}                 
 \def\PXi         {\ensuremath{\Xi}\xspace}                 
 \def\PLambda     {\ensuremath{\Lambda}\xspace}                 
 \def\PSigma      {\ensuremath{\Sigma}\xspace}                 
 \def\POmega      {\ensuremath{\Omega}\xspace}                 
 \def\PUpsilon    {\ensuremath{\Upsilon}\xspace}

 \def\PB      {\ensuremath{\mathrm{B}}\xspace}                 
                  
 \def\PD      {\ensuremath{\mathrm{D}}\xspace}

 \def\PJ      {\ensuremath{\mathrm{J}}\xspace}                 
 \def\PK      {\ensuremath{\mathrm{K}}\xspace}

 \def\Pb      {\ensuremath{\mathrm{b}}\xspace}                 
 \def\Pc      {\ensuremath{\mathrm{c}}\xspace}

 \def\Pi      {\ensuremath{\mathrm{i}}\xspace}

 \def\Ps      {\ensuremath{\mathrm{s}}\xspace}

 \def\thebaroffset{0.0em}
}
{

 \def\Pmu         {\ensuremath{\mu}\xspace}                 
 \def\Pnu         {\ensuremath{\nu}\xspace}                 
                  
 \def\Ppi         {\ensuremath{\pi}\xspace}

 \def\Ppsi        {\ensuremath{\psi}\xspace}                 
                  
 \mathchardef\PDelta="7101
 \mathchardef\PXi="7104
 \mathchardef\PLambda="7103
 \mathchardef\PSigma="7106
 \mathchardef\POmega="710A
 \mathchardef\PUpsilon="7107
                  
 \def\PB      {\ensuremath{B}\xspace}                 
                  
 \def\PD      {\ensuremath{D}\xspace}

 \def\PJ      {\ensuremath{J}\xspace}                 
 \def\PK      {\ensuremath{K}\xspace}

 \def\Pb      {\ensuremath{b}\xspace}                 
 \def\Pc      {\ensuremath{c}\xspace}

 \def\Pi      {\ensuremath{i}\xspace}

 \def\Ps      {\ensuremath{s}\xspace}

 \def\thebaroffset{0.18em}
}
\newcommand{\offsetoverline}[2][\thebaroffset]{\kern #1\overline{\kern -#1 #2}}%

\makeatletter
\ifcase \@ptsize \relax
  \newcommand{\miniscule}{\@setfontsize\miniscule{4}{5}}
\or
  \newcommand{\miniscule}{\@setfontsize\miniscule{5}{6}}
\or
  \newcommand{\miniscule}{\@setfontsize\miniscule{5}{6}}
\fi
\makeatother

\DeclareRobustCommand{\optbar}[1]{\shortstack{{\miniscule (\rule[.5ex]{1.25em}{.18mm})}
  \\ [-.7ex] $#1$}}

\def\mup        {{\ensuremath{\Pmu^+}}\xspace}
\def\mun        {{\ensuremath{\Pmu^-}}\xspace} 

\def\neub       {{\ensuremath{\overline{\Pnu}}}\xspace}

\def\neumb      {{\ensuremath{\neub_\mu}}\xspace}

\def\squark    {{\ensuremath{\Ps}}\xspace}

\def\cquark    {{\ensuremath{\Pc}}\xspace}

\def\bquark    {{\ensuremath{\Pb}}\xspace}
\def\bquarkbar {{\ensuremath{\overline \bquark}}\xspace}
\def\bbbar     {{\ensuremath{\bquark\bquarkbar}}\xspace}

\def\pion   {{\ensuremath{\Ppi}}\xspace}

\def\pip    {{\ensuremath{\pion^+}}\xspace}
\def\pim    {{\ensuremath{\pion^-}}\xspace}

\def\kaon    {{\ensuremath{\PK}}\xspace}

\def\KorKbar {\kern \thebaroffset\optbar{\kern -\thebaroffset \PK}{}\xspace}

\def\Kp      {{\ensuremath{\kaon^+}}\xspace}
\def\Km      {{\ensuremath{\kaon^-}}\xspace}

\def\D       {{\ensuremath{\PD}}\xspace}

\def\DorDbar {\kern \thebaroffset\optbar{\kern -\thebaroffset \PD}\xspace}

\def\Dm      {{\ensuremath{\D^-}}\xspace}

\def\Dsm     {{\ensuremath{\D^-_\squark}}\xspace}

\def\B       {{\ensuremath{\PB}}\xspace}

\def\BorBbar {\kern \thebaroffset\optbar{\kern -\thebaroffset \PB}\xspace}
\def\Bz      {{\ensuremath{\B^0}}\xspace}

\def\Bub     {{\ensuremath{\B^-}}\xspace}

\def\Bm      {{\ensuremath{\Bub}}\xspace}

\def\Bs      {{\ensuremath{\B^0_\squark}}\xspace}

\def\jpsi     {{\ensuremath{{\PJ\mskip -3mu/\mskip -2mu\Ppsi\mskip 2mu}}}\xspace}

\def\Y#1S{\ensuremath{\PUpsilon{(#1S)}}\xspace}

\def\Lz          {{\ensuremath{\PLambda}}\xspace}
\def\Lbar        {{\ensuremath{\offsetoverline{\PLambda}}}\xspace}
\def\LorLbar     {\kern \thebaroffset\optbar{\kern -\thebaroffset \PLambda}\xspace}

\def\Xires       {{\ensuremath{\PXi}}\xspace}

\def\Xiresm      {{\ensuremath{\Xires^-}}\xspace}
\def\Xiresbar    {{\ensuremath{\offsetoverline{\Xires}}}\xspace}

\def\Xiresbarp   {{\ensuremath{\Xiresbar^+}}\xspace}
\def\Omegares    {{\ensuremath{\POmega}}\xspace}

\def\Lc          {{\ensuremath{\Lz^+_\cquark}}\xspace}

\def\Xicz        {{\ensuremath{\Xires^0_\cquark}}\xspace}
\def\Xicp        {{\ensuremath{\Xires^+_\cquark}}\xspace}

\def\Lb           {{\ensuremath{\Lz^0_\bquark}}\xspace}
\def\Lbbar        {{\ensuremath{\Lbar{}^0_\bquark}}\xspace}

\def\Xib          {{\ensuremath{\Xires_\bquark}}\xspace}
\def\Xibz         {{\ensuremath{\Xires^0_\bquark}}\xspace}
\def\Xibm         {{\ensuremath{\Xires^-_\bquark}}\xspace}

\def\Xibbarp      {{\ensuremath{\Xiresbar{}_\bquark^+}}\xspace}
\def\Omegab       {{\ensuremath{\Omegares^-_\bquark}}\xspace}

\def\BF         {{\ensuremath{\mathcal{B}}}\xspace}
\def\BR         {\BF}

\def\to                 {\ensuremath{\rightarrow}\xspace}

\def\CP                {{\ensuremath{C\!P}}\xspace}

\def\AT#1     {\ensuremath{A_{\mathrm{T}}^{#1}}\xspace}           

\def\C#1      {\ensuremath{\mathcal{C}_{#1}}\xspace}                       
\def\Cp#1     {\ensuremath{\mathcal{C}_{#1}^{'}}\xspace}                    
\def\Ceff#1   {\ensuremath{\mathcal{C}_{#1}^{\mathrm{(eff)}}}\xspace}        
\def\Cpeff#1  {\ensuremath{\mathcal{C}_{#1}^{'\mathrm{(eff)}}}\xspace}       
\def\Ope#1    {\ensuremath{\mathcal{O}_{#1}}\xspace}                       
\def\Opep#1   {\ensuremath{\mathcal{O}_{#1}^{'}}\xspace}                    

\newcommand{\nospaceunit}[1]{\ensuremath{\text{#1}}}       
\newcommand{\aunit}[1]{\ensuremath{\text{\,#1}}}       

\newcommand{\tev}{\aunit{Te\kern -0.1em V}\xspace}
\newcommand{\gev}{\aunit{Ge\kern -0.1em V}\xspace}
\newcommand{\mev}{\aunit{Me\kern -0.1em V}\xspace}
\newcommand{\kev}{\aunit{ke\kern -0.1em V}\xspace}
\newcommand{\ev}{\aunit{e\kern -0.1em V}\xspace}
\newcommand{\mevc}{\ensuremath{\aunit{Me\kern -0.1em V\!/}c}\xspace}
\newcommand{\gevc}{\ensuremath{\aunit{Ge\kern -0.1em V\!/}c}\xspace}
\newcommand{\mevcc}{\ensuremath{\aunit{Me\kern -0.1em V\!/}c^2}\xspace}
\newcommand{\gevcc}{\ensuremath{\aunit{Ge\kern -0.1em V\!/}c^2}\xspace}

\def\mum  {\ensuremath{\,\upmu\nospaceunit{m}}\xspace}

\def\fb   {\ensuremath{\aunit{fb}}\xspace}
\def\invfb   {\ensuremath{\fb^{-1}}\xspace}

\newcommand{\chisq}{\ensuremath{\chi^2}\xspace}

\newcommand{\chisqip}{\ensuremath{\chi^2_{\text{IP}}}\xspace}

\def\gsim{{~\raise.15em\hbox{$>$}\kern-.85em
          \lower.35em\hbox{$\sim$}~}\xspace}
\def\lsim{{~\raise.15em\hbox{$<$}\kern-.85em
          \lower.35em\hbox{$\sim$}~}\xspace}

\def\sPlot{\mbox{\em sPlot}\xspace}

\def\pt         {\ensuremath{p_{\mathrm{T}}}\xspace}

\def\ptot       {\ensuremath{p}\xspace}

\def\evtgen     {\mbox{\textsc{EvtGen}}\xspace}

\def\geant      {\mbox{\textsc{Geant4}}\xspace}

\def\photos     {\mbox{\textsc{Photos}}\xspace}

\def\pythia     {\mbox{\textsc{Pythia}}\xspace}

\xspace

\def\tell1  {TELL1\xspace}
\def\ukl1   {UKL1\xspace}

\usepackage{cite} 
\usepackage{mciteplus}

\usepackage{longtable} 

\begin{document}

\renewcommand{\thefootnote}{\fnsymbol{footnote}}
\setcounter{footnote}{1}


\begin{titlepage}
\pagenumbering{roman}

\vspace*{-1.5cm}
\centerline{\large EUROPEAN ORGANIZATION FOR NUCLEAR RESEARCH (CERN)}
\vspace*{1.0cm}
\noindent
\begin{tabular*}{\linewidth}{lc@{\extracolsep{\fill}}r@{\extracolsep{0pt}}}
\ifthenelse{\boolean{pdflatex}}
{\vspace*{-1.5cm}\mbox{\!\!\!\includegraphics[width=.14\textwidth]{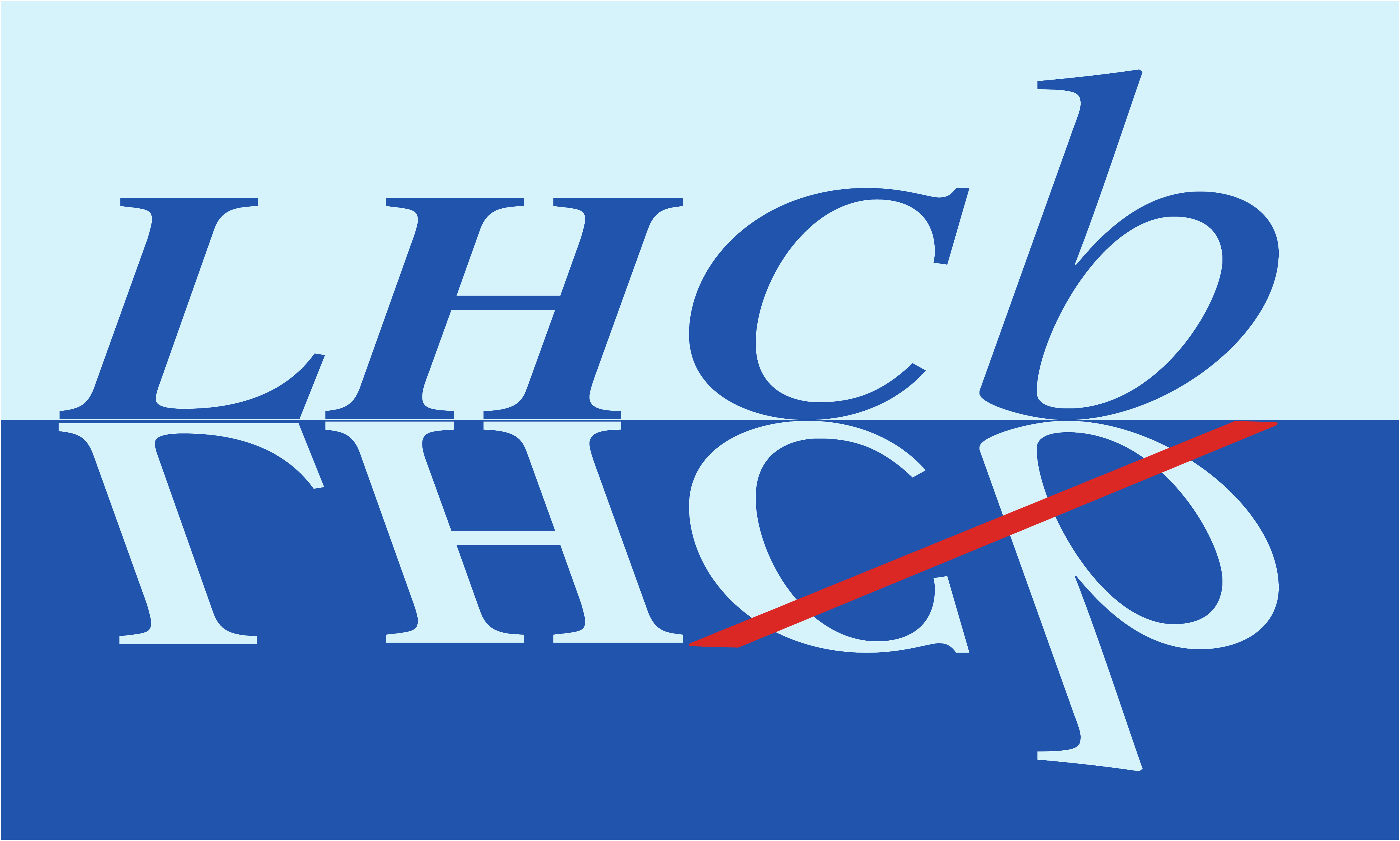}} & &}%
{\vspace*{-1.2cm}\mbox{\!\!\!\includegraphics[width=.12\textwidth]{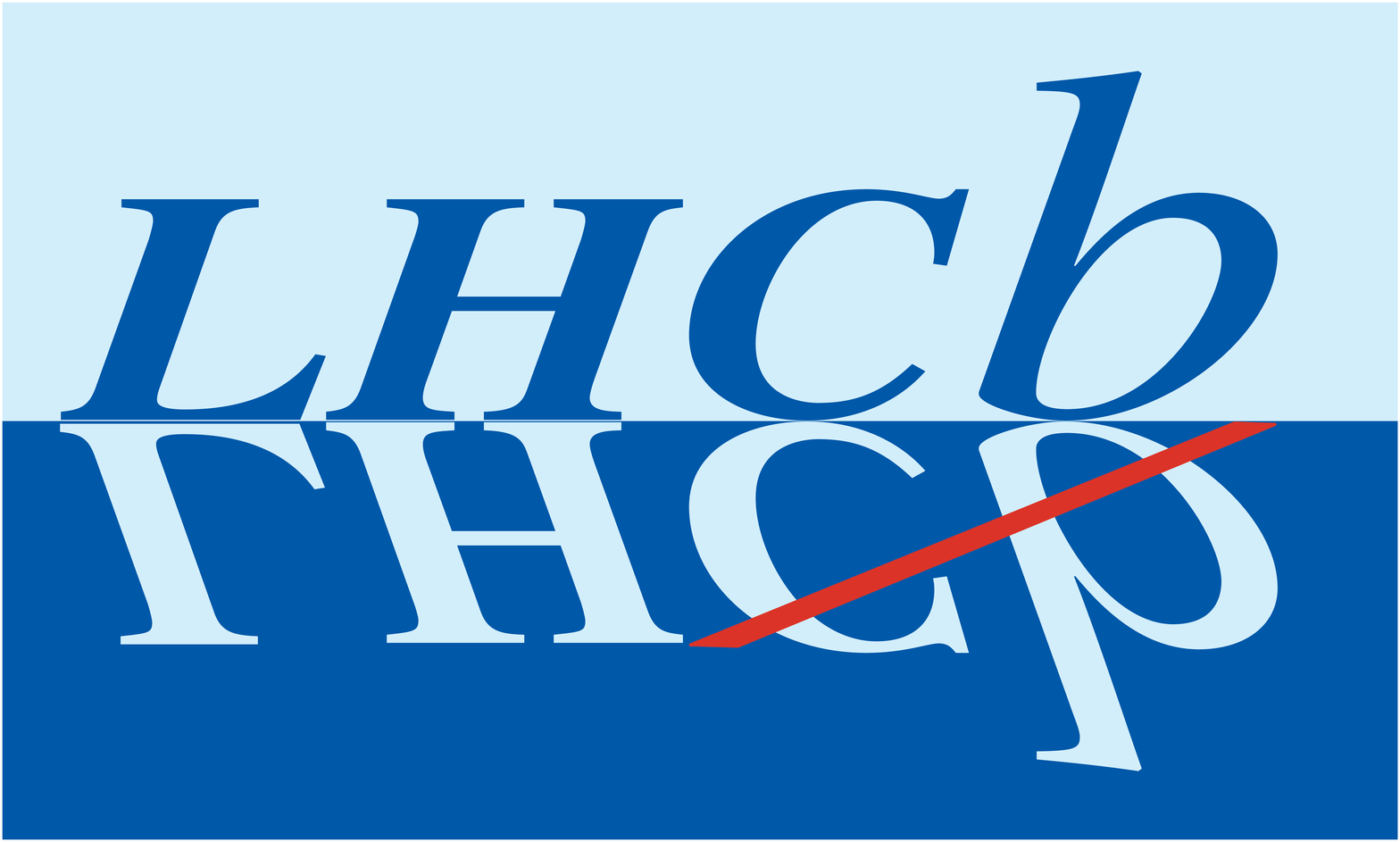}} & &}%
\\
 & & CERN-EP-2018-348 \\  
 & & LHCb-PAPER-2018-047 \\  
 & & \today \\ 
 & & \\
\end{tabular*}

\vspace*{1.0cm}

{\normalfont\bfseries\boldmath\huge
\begin{center}
  \papertitle 
\end{center}
}

\vspace*{1.0cm}

\begin{center}
\paperauthors\footnote{Authors are listed at the end of this paper.}
\end{center}

\vspace{\fill}

\begin{abstract}
  \noindent
The first measurement of the production rate of $\Xibm$ baryons in $pp$ collisions relative to that 
of $\Lb$ baryons is reported, using data samples collected by the LHCb experiment, and
corresponding to integrated luminosities of 1, 2 and 1.6\invfb at $\sqrt{s}=7,~8$ and 13\tev, respectively.
In the kinematic region  $2 < \eta < 6$ and $\pt<20$\gevc, we measure
\begin{align*}
\frac{f_{\Xibm}}{f_{\Lb}}\frac{\BR(\Xibm\to\jpsi\Xiresm)}{\BR(\Lb\to\jpsi\Lz)} &=  (10.8\pm0.9\pm0.8)\times10^{-2}~~[\sqrt{s}=7,8\tev],\\
\frac{f_{\Xibm}}{f_{\Lb}}\frac{\BR(\Xibm\to\jpsi\Xiresm)}{\BR(\Lb\to\jpsi\Lz)} &= (13.1\pm1.1\pm1.0)\times10^{-2}~~[\sqrt{s}=13\tev],
\end{align*}
where $f_{\Xibm}$ and $f_{\Lb}$ are the fragmentation fractions of $b$ quarks into $\Xibm$ and $\Lb$ baryons,
respectively, $\BR$ represents branching fractions, and the uncertainties are due to statistical and experimental 
systematic sources. The values of $f_{\Xibm}/f_{\Lb}$ are obtained by invoking SU(3) symmetry in the $\Xibm\to\jpsi\Xiresm$ 
and $\Lb\to\jpsi\Lz$ decays. Production asymmetries between $\Xibm$ and $\Xibbarp$ baryons are also reported.

The mass of the $\Xibm$ baryon is also measured relative to that of the $\Lb$ baryon, from which it is 
found that
\begin{align*}
m(\Xibm) &= 5796.70\pm0.39\pm0.15\pm0.17\mevcc,
\end{align*}
\noindent where the last uncertainty is due to the precision on the known $\Lb$ mass.
This result represents the most precise determination of the $\Xibm$ mass.

\end{abstract}

\vspace*{1.0cm}

\begin{center}
  Published in Phys. Rev. D99 052006 (2019)
\end{center}

\vspace{\fill}

{\footnotesize 
\centerline{\copyright~\papercopyright. \href{\paperlicenceurl}{\paperlicence}.}}
\vspace*{2mm}

\end{titlepage}


\newpage
\setcounter{page}{2}
\mbox{~}
%
%
%
%

\cleardoublepage


\renewcommand{\thefootnote}{\arabic{footnote}}
\setcounter{footnote}{0}



\pagestyle{plain} 
\setcounter{page}{1}
\pagenumbering{arabic}


%

The decays of beauty ($b$) quarks provide a sensitive probe of physics within, and beyond, the Standard Model. 
Due to the large $\bbbar$ production cross-section at the Large Hadron Collider, beauty hadrons of all species are abundantly
produced.  Measurements of branching fractions in specific decay channels are often needed in order to make
quantitative comparisons to theoretical predictions.
However, absolute branching fraction measurements at hadron colliders are difficult
to perform without external input. Instead, one generally resorts to measuring a particular
branching fraction relative to that of a topologically similar decay mode, frequently one that involves either a $\Bz$ or a $\Bm$ meson,
whose absolute branching fractions are known from $B$-factory measurements.
When $\Bs$ or $\Lb$ branching fractions are measured relative to those of a $\Bz$ decay,
knowledge of the ratio of fragmentation fractions, $f_s/f_d$ for $\Bs$ decays, or $f_{\Lb}/f_d$ for $\Lb$ decays, is required.
Here, $f_d$, $f_s$ and $f_{\Lb}$ represents the rates at which a $b$ quark
hadronizes into a $\Bz$, $\Bs$ or $\Lb$ hadron, respectively.

Theoretically, the most robust way to measure the $b$-quark fragmentation fractions is to exploit the well tested
prediction from heavy quark effective 
theory~\cite{Khoze:1983yp,Bigi:1991ir,Bigi:1992su,Blok:1992hw,Blok:1992he,Neubert:1997gu,Uraltsev:1998bk,Bellini:1996ra} 
that, to first order, all $b$ hadrons containing a single heavy quark have equal
semileptonic decay widths.  Such analyses have been carried out by the LHCb experiment at $\sqrt{s}=7\tev$~\cite{LHCb-PAPER-2011-018}
and 13\tev~\cite{LHCb-PAPER-2018-050}, where it is found that $\langle f_s/f_d\rangle\simeq0.26$ and 
$\langle f_{\Lb}/f_d\rangle\simeq0.6$, 
averaged over the pseudorapidity ($\eta$) and transverse momentum ($\pt$) region 
$2<\eta<5$ and $3<\pt<25$\gevc. An alternative technique, which relies on  
factorization and SU(3) flavor symmetry in the 
$\Bs\to\Dsm\pip$ and $\Bz\to\Dm\Kp$ decays~\cite{LHCb-PAPER-2012-037}, has also been used to measure 
$f_s/f_d$, yielding a value consistent with that obtained in semileptonic decays. 

With the large samples of $b$ hadrons collected by the LHCb experiment, a number of new decay modes of 
$\Xibz$, $\Xibm$, and even $\Omegab$ baryons have been searched for, and in many cases have
led to first 
observations~\cite{LHCb-PAPER-2013-056,LHCb-PAPER-2013-061,LHCb-PAPER-2014-021,LHCb-PAPER-2014-048,LHCb-PAPER-2015-047,LHCb-PAPER-2016-004,LHCb-PAPER-2016-050,LHCb-PAPER-2016-053,LHCb-PAPER-2017-034}.
However, when new decay modes of these baryons are observed, 
absolute branching fractions cannot be determined due to a lack of knowledge of the fragmentation
fractions $f_{\Xibz}$, $f_{\Xibm}$ and $f_{\Omegab}$. For example, in one such measurement, evidence of the strangeness-changing weak
decay $\Xibm\to\Lb\pim$ is reported~\cite{LHCb-PAPER-2015-047}, with the result that
$(f_{\Xibm}/f_{\Lb})\BR(\Xibm\to\Lb\pim)=(5.7\pm1.8^{+0.8}_{-0.9})\times10^{-4}$. To compute the decay width
$\Gamma(\Xibm\to\Lb\pim)$ and compare to theoretical predictions requires knowledge of the ratio $f_{\Xibm}/f_{\Lb}$.

In principle, the same procedure used to measure $f_s/f_d$ and $f_{\Lb}/f_d$ can be applied to semileptonic 
$\Xibz\to\Xicp\mun\neumb X$ and $\Xibm\to\Xicz\mun\neumb X$ decays to measure $f_{\Xibz}/f_d$ and $f_{\Xibm}/f_d$. 
However, an obstacle to such an analysis is the 
limited knowledge of absolute branching fractions for the decays of the $\Xicp$ or $\Xicz$ baryon. Recently,
the Belle experiment published a first measurement of the absolute branching fractions for three $\Xicz$ decay modes, each
with a relative precision of about 40\%~\cite{Li:2018qak}. No such measurements exist yet for the $\Xicp$ baryon. Precise measurements of 
branching fractions for both $\Xicp$ and $\Xicz$ decays should be feasible in the Belle II experiment~\cite{Abe:2010gxa}.

Production ratio measurements of the hadronic $\Xibz\to\Xicp\pim$ and $\Lb\to\Lc\pim$ decays~\cite{LHCb-PAPER-2014-021},
where both the $\Xicp$ and $\Lc$ baryons are reconstructed in the $p\Km\pip$ final state, have been used to
predict $f_{\Xibz}/f_{\Lb}$. In this case, theoretical estimates of $\BR(\Xicp\to p\Km\pip)$ are used, resulting in 
predictions of $f_{\Xibz}/f_{\Lb}=0.065\pm0.020$~\cite{DiWang} and $f_{\Xibz}/f_{\Lb}=0.054\pm0.020$~\cite{Jiang:2018iqa}.

An alternative approach to either of these two methods is to exploit the decays $\Lb\to\jpsi\Lz$ and $\Xibm\to\jpsi\Xiresm$, where
the $\Xiresm$ baryon is detected in its decay to $\Lz\pim$.  Charge-conjugate processes are implicitly included. 
These decay rates are related through SU(3) flavor symmetry, where one finds~\cite{SAVAGE198915,voloshin,Hsiao:2015txa}
\begin{align}
\frac{\Gamma(\Xibm\to\jpsi\Xiresm)}{\Gamma(\Lb\to\jpsi\Lz)} = \frac{3}{2}.
\label{eq:suthree}
\end{align}
The ratio
\begin{align}
R\equiv\frac{f_{\Xibm}}{f_{\Lb}}\frac{\BR(\Xibm\to\jpsi\Xiresm)}{\BR(\Lb\to\jpsi\Lz)}=\frac{f_{\Xibm}}{f_{\Lb}}\frac{\Gamma(\Xibm\to\jpsi\Xiresm)}{\Gamma(\Lb\to\jpsi\Lz)}\frac{\tau_{\Xibm}}{\tau_{\Lb}}
\label{eq:defR}
\end{align}
\noindent depends on $f_{\Xibm}/f_{\Lb}$, the partial decay widths, $\Gamma$, and the lifetimes, $\tau$, of the indicated $b$ baryons.
Experimentally, $R$ is obtained from the ratio of efficiency-corrected yields
\begin{align}
R = \frac{N(\Xibm\to\jpsi\Xiresm)}{N(\Lb\to\jpsi\Lz)}\frac{\epsilon_{\Lb}}{\epsilon_{\Xibm}},
\label{eq:Req}
\end{align}
\noindent where $\epsilon$ represents the detection efficiency and $N$ is the yield of the indicated decays.

In this article, we report a first measurement of the ratio $R$ in $pp$ collision data collected by the LHCb experiment, 
corresponding to integrated luminosities of 1.0\invfb at $\sqrt{s}=7$\tev, 2.0\invfb at $\sqrt{s}=8$\tev and
1.6\invfb at $\sqrt{s}=13$\tev. The measurement of $R$, along with the SU(3) assumption in Eq.~\ref{eq:suthree} and
the known $\Lb$ and $\Xibm$ baryon lifetimes~\cite{PDG2018}, is used to infer 
the value of $f_{\Xibm}/f_{\Lb}$. The same data samples are also used to measure
the production asymmetry between $\Xibm$ and $\Xibbarp$ baryons, and
make the most precise measurement of the $\Xibm$ mass.

The \lhcb detector~\cite{Alves:2008zz,LHCb-DP-2014-002} is a single-arm forward
spectrometer designed for the study of particles containing \bquark or \cquark
quarks. The detector includes a high-precision tracking system
consisting of a silicon-strip vertex detector surrounding the $pp$
interaction region, a large-area silicon-strip detector located
upstream of a dipole magnet with a bending power of about
$4{\mathrm{\,Tm}}$, and three stations of silicon-strip detectors and straw
drift tubes placed downstream of the magnet.
The tracking system provides a measurement of the momentum, \ptot, of charged particles with
a relative uncertainty that varies from 0.5\% at low momentum to 1.0\% at 200\gevc.
The minimum distance of a track to a primary vertex (PV), the impact parameter (IP), 
is measured with a resolution of $(15+29/\pt)\mum$,
where \pt is expressed in\,\gevc.
Different types of charged hadrons are distinguished using information
from two ring-imaging Cherenkov detectors.
Photons, electrons and hadrons are identified by a calorimeter system consisting of
scintillating-pad and preshower detectors, an electromagnetic
and a hadronic calorimeter. Muons are identified by a
system composed of alternating layers of iron and multiwire
proportional chambers. The online event selection is performed by a trigger
which consists of a hardware stage, based on information from the calorimeter and muon
systems, followed by a software stage, which applies a full event
reconstruction.

Simulation is required to model the effects of the detector acceptance and the
imposed selection requirements. In the simulation, $pp$ collisions are generated using
\pythia~\cite{Sjostrand:2006za,*Sjostrand:2007gs} with a specific \lhcb
configuration~\cite{LHCb-PROC-2010-056}.  Decays of unstable particles
are described by \evtgen~\cite{Lange:2001uf}, in which final-state
radiation is generated using \photos~\cite{Golonka:2005pn}. The
interaction of the generated particles with the detector, and its response,
are implemented using the \geant toolkit~\cite{Allison:2006ve, *Agostinelli:2002hh} as described in
Ref.~\cite{LHCb-PROC-2011-006}.

The $\Xibm\to\jpsi\Xiresm(\to\Lz\pim)$ and $\Lb\to\jpsi\Lz$ decays both contain a $\jpsi$ meson and a $\Lz$ baryon in the decay chain,
and are kinematically similar. To reduce systematic uncertainties,
selection requirements are tailored to exploit the common particles in the final state of the $\Lb$ and $\Xibm$ decays.
At the trigger level, both modes are required to satisfy requirements based solely on the $\jpsi\to\mup\mun$ decay.
Firstly, the hardware stage must register either a single high-$\pt$ muon
or a $\mup\mun$ pair. The software stage~\cite{LHCb-DP-2012-004} then requires a $\mup\mun$ pair whose
decay vertex is displaced from all PVs in the event, and 
that has an invariant mass consistent with the known $\jpsi$ mass~\cite{PDG2018}.

Selected events may contain more than one PV. Each particle is associated
to the PV for which the corresponding value of $\chisqip$ is smallest, where
\chisqip\ is defined as the difference in the vertex-fit \chisq of a given PV reconstructed with and
without the particle under consideration.

In the offline analysis, each muon must have $\pt$ in excess of 550$\mevc$ and have IP
to all PVs in the event that exceeds approximately three times the expected uncertainty.
The $\mup\mun$ pair must form a good-quality vertex 
and have an invariant mass within 40\mevcc of the known $\jpsi$ mass, corresponding to about three 
times the mass resolution.

Reconstructed charged particles are classified into two categories in this analysis.
The {\emph{long}} category refers to tracks that have reconstructed segments in both the 
vertex detector and the tracking stations upstream and downstream of the LHCb magnet.
The {\emph{downstream}} category consists of those tracks that are not reconstructed in
the vertex detector, and thus only include information from the tracking detectors just before
and after the LHCb magnet. While most of the reconstructed particles from the $pp$ interactions are
in the long category, the decay products of long-lived strange particles tend to be mostly reconstructed as
downstream tracks. Because of the presence of vertex detector measurements, the
trajectories, and hence the IP, of long tracks are measured with better precision than those of
downstream tracks.

Candidate $\Lz\to p\pim$ decays are formed by combining downstream $p$ and $\pim$ candidates with
$\pt$ in excess of 500\mevc and 100\mevc, respectively. Both tracks
are required to be significantly detached from all PVs in the event, and together they must form a good-quality 
vertex and must satisfy the requirement $|M(p\pim)-m_{\Lz}|<8$\mevcc, corresponding to approximately three 
times the mass resolution. Here and throughout the text, $M$ represents 
an invariant mass and $m$ represents the known mass of the indicated particle~\cite{PDG2018}. 

The $\Xiresm$ baryon is reconstructed through its decay to $\Lz\pim$.
Due to the long $\Xiresm$ and $\Lz$ lifetimes, only $\Lz$ candidates formed from downstream tracks are used,
as they contribute about 90\% to the $\Xiresm$ sample in $\Xibm$ decays. To maintain a uniform selection,
the same requirement is imposed on $\Lz$ decays in the $\Lb$ mode. The
$\pim$ meson from the $\Xiresm$ decay may be reconstructed as either a long or a downstream track.  
For the $\Xibm$ mass and production asymmetry measurements, both categories
are used. However, for the measurement of $R$, only the long-track sample is used, since
the efficiency for detecting the $\pim$ meson in the decay $\Xiresm\to\Lz\pim$ enters directly in Eq.~\ref{eq:Req}, and long-track 
efficiencies have been precisely calibrated using a tag-and-probe method~\cite{LHCb-DP-2013-002}. No explicit momentum requirement
is applied to the $\pim$ meson, since it typically has low momentum. When necessary, the notation
$\pi^-_{\rm L}$ and $\pi^-_{\rm D}$ is used to distinguish between long~(L) and downstream~(D) $\pim$ tracks.
Tracks in the $\pi^-_{\rm L}$ sample are required to be significantly detached from all PVs in the event,
corresponding to a requirement that the impact parameter exceeds about four times the corresponding uncertainty;
no such requirement is necessary on
the $\pi^-_{\rm D}$ sample. Exploiting the large $\Xiresm$ baryon lifetime, $\Xiresm$ candidates
must have $t_{\rm PV}>6$~ps, where $t_{\rm PV}$ is the decay time measured relative to the associated PV.
Lastly, $\Xiresm$ candidates are required to satisfy the mass requirement 
$|M(\Lz\pi^{-}_{\rm L,D})-M(p\pim)+m_{\Lz}-m_{\Xiresm}|<10$\mevcc, corresponding to about three times the mass resolution, 
and have positive decay time, measured relative to the $\Xibm$ decay vertex. 

The $\Lb$ ($\Xibm$) candidates are formed by combining $\jpsi$ and $\Lz$ ($\Xiresm$) candidates. A vertex fit of good quality
is required. To suppress background from prompt $\jpsi$ production,
the $b$ hadron is required to have a reconstructed decay time larger than 0.2~ps, which is about four times
the resolution. Finally, to have
a well-defined fiducial region, the $\Lb$ and $\Xibm$ candidates are required to be within the kinematic region
$2<\eta<6$ and $\pt<20$\gevc. Multiple candidates in a single event occur in less than 1\% of selected events, and all
candidates are kept. To improve the mass resolution, an
additional kinematic fit is performed on each candidate, employing both vertex and mass constraints on
the $\jpsi$, $\Lz$ and $\Xiresm$ candidates~\cite{Hulsbergen:2005pu}. 
The resulting mass resolution is about 8\mevcc for both modes.

\begin{figure}[p]
\centering
\includegraphics[width=0.44\textwidth]{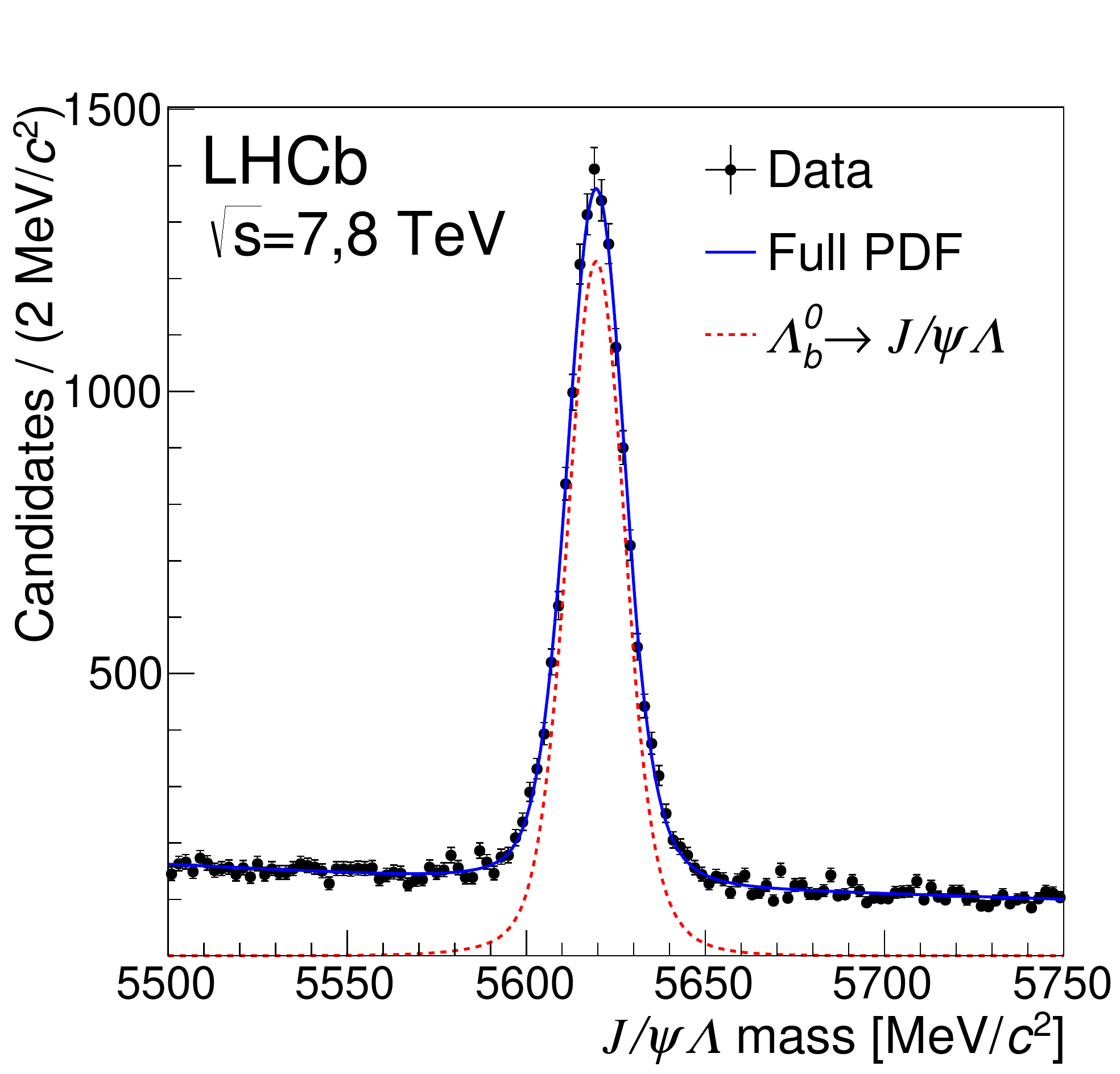}
\includegraphics[width=0.44\textwidth]{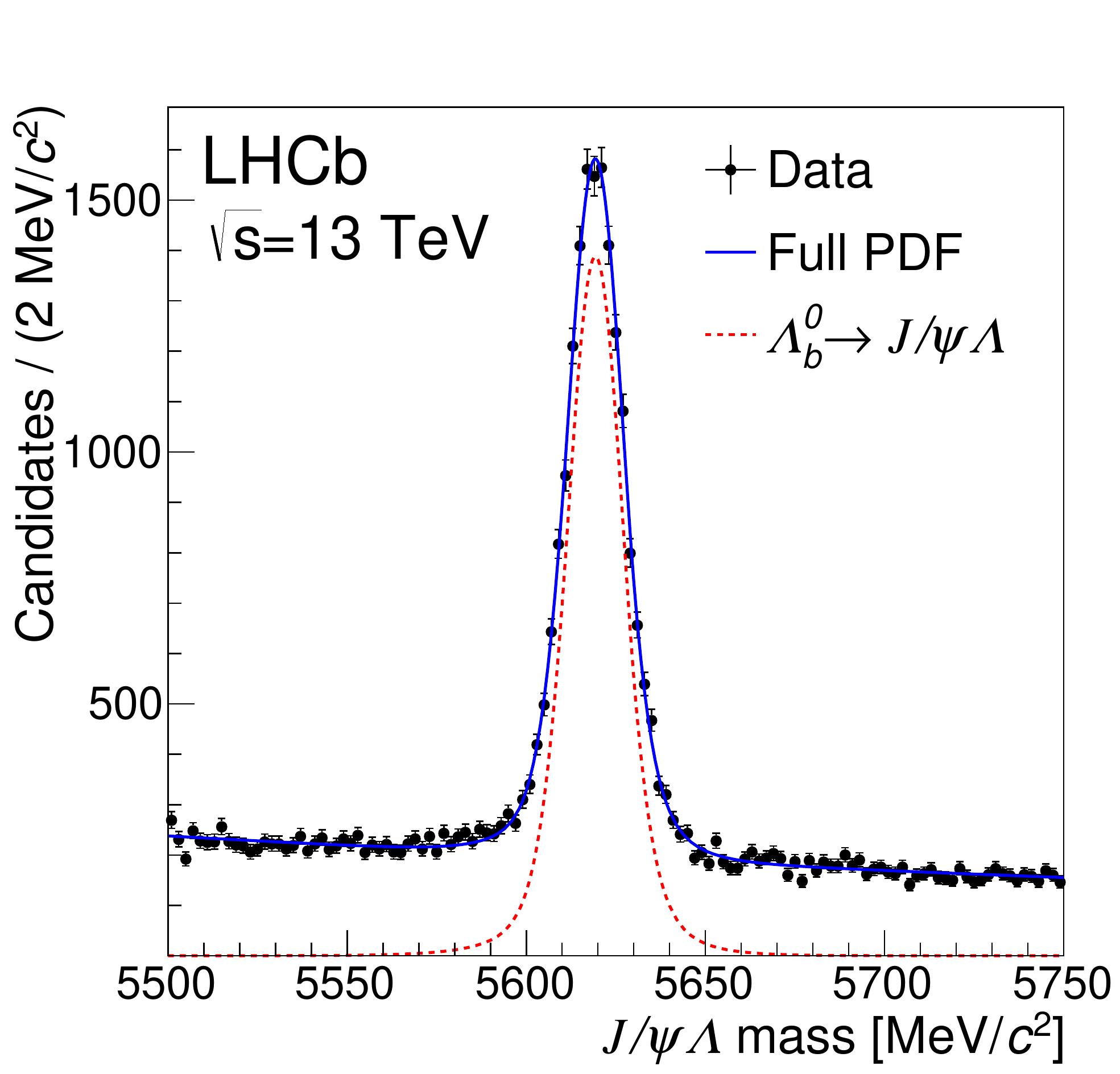}
\includegraphics[width=0.44\textwidth]{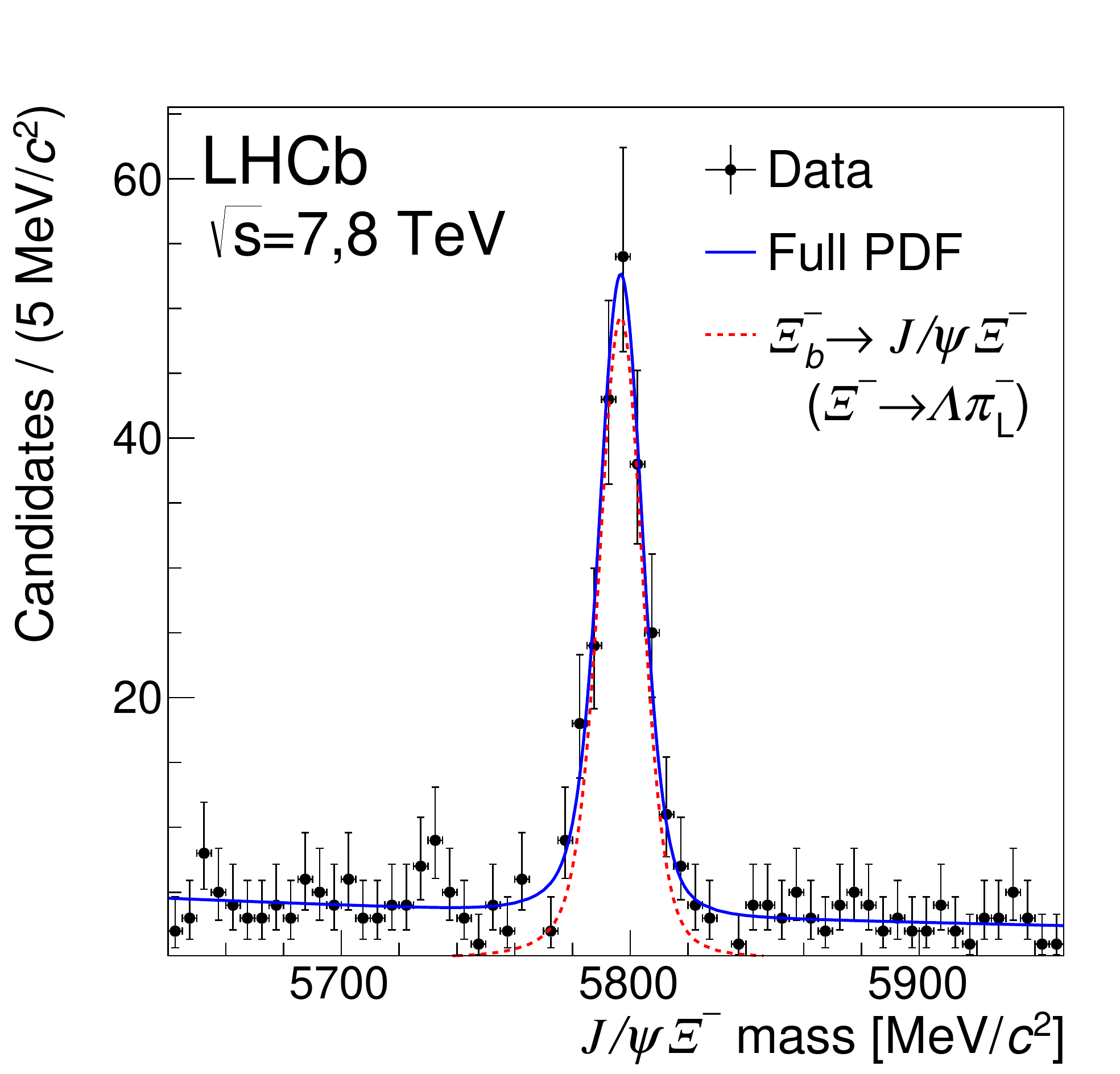}
\includegraphics[width=0.44\textwidth]{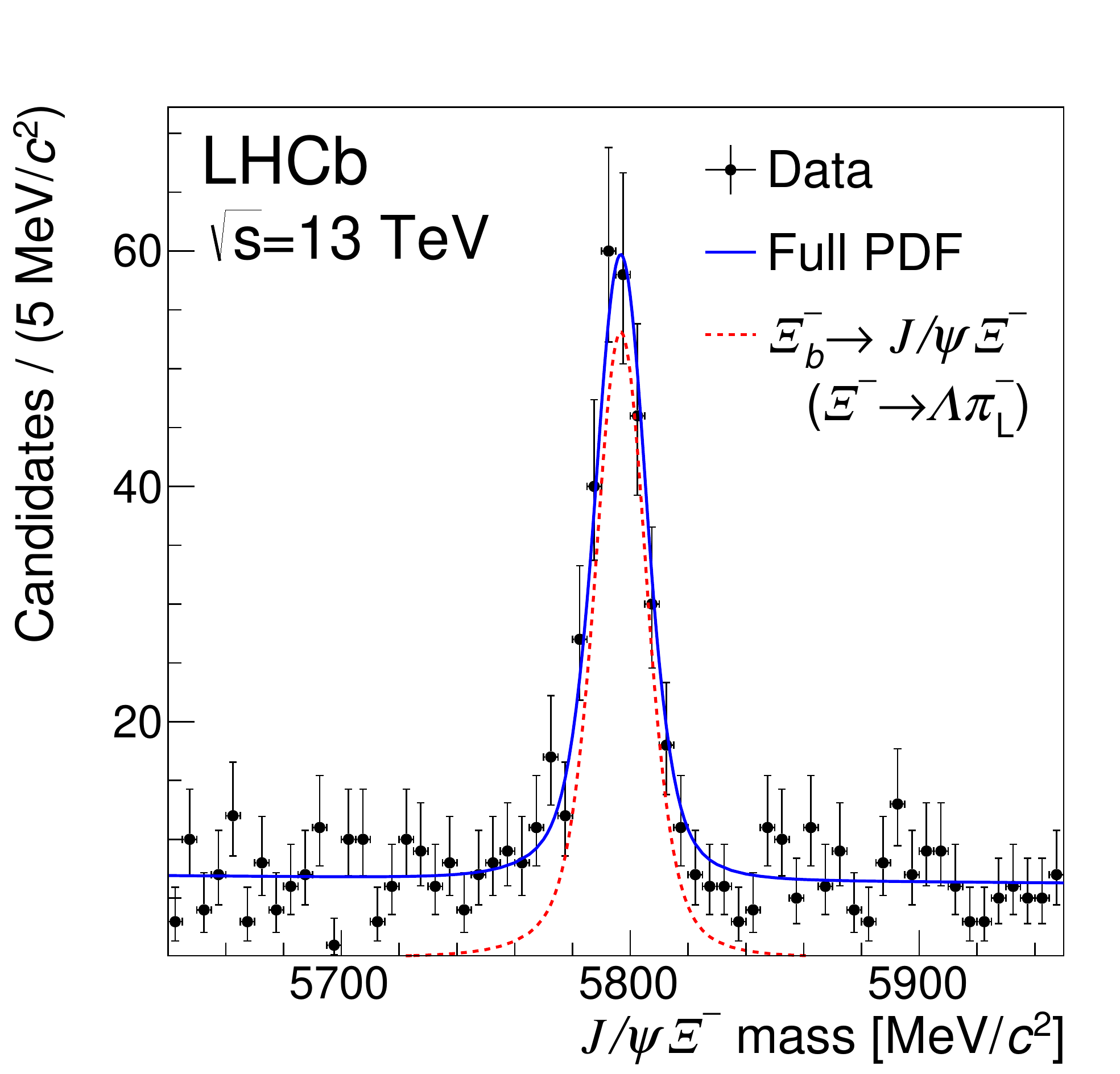}
\includegraphics[width=0.44\textwidth]{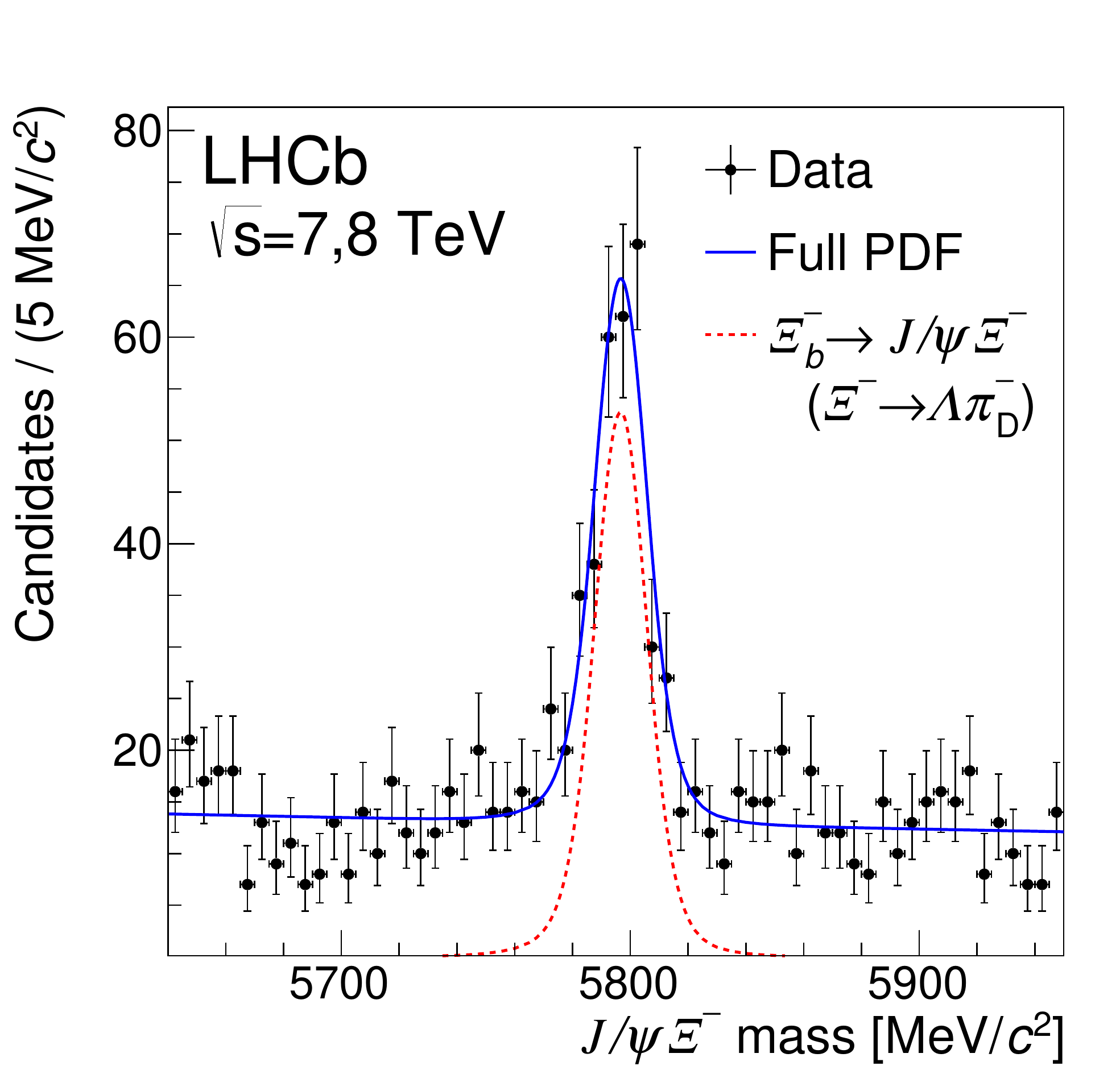}
\includegraphics[width=0.44\textwidth]{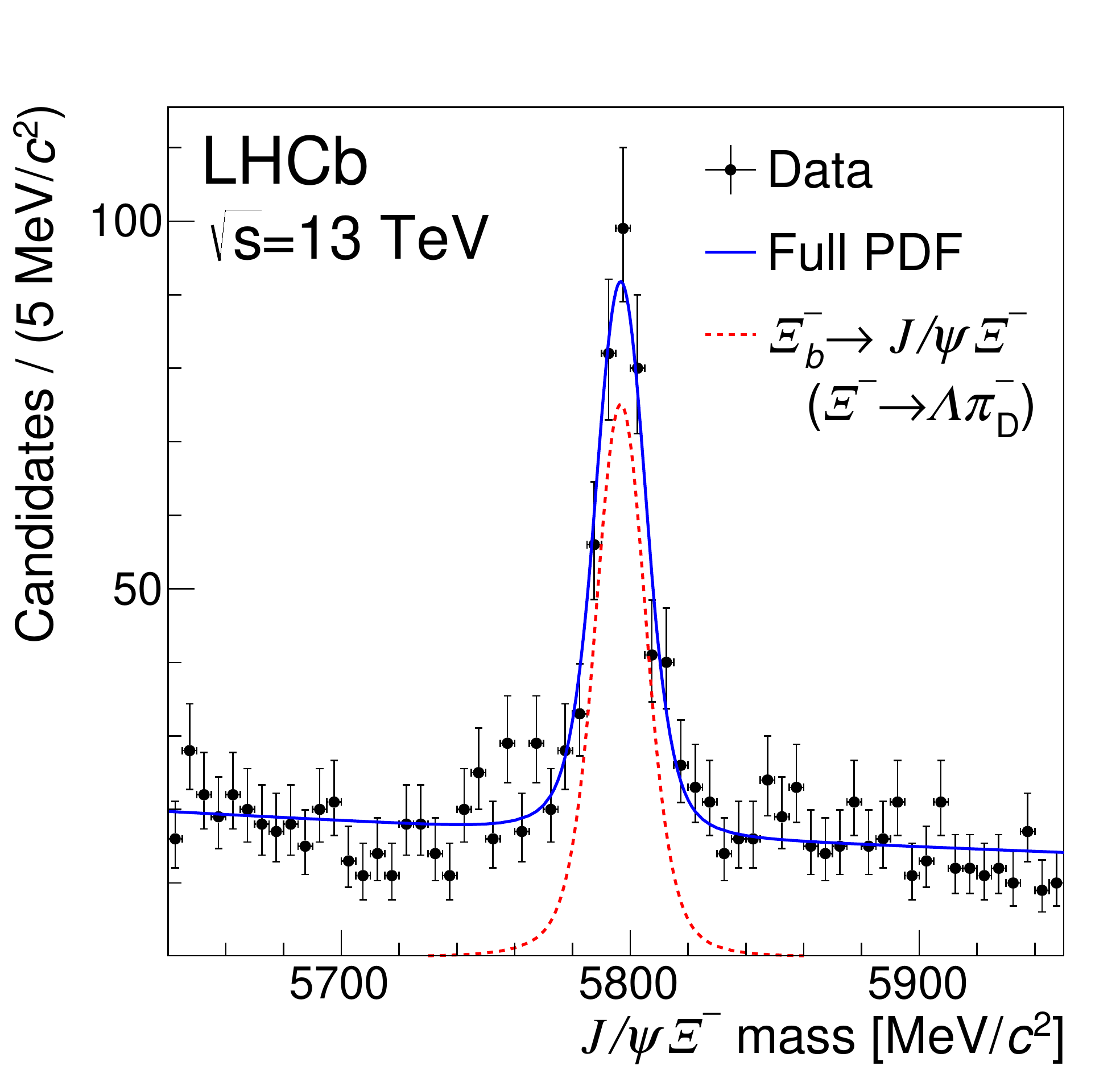}
\caption{\small{Invariant-mass distributions for (top) $\Lb\to\jpsi\Lz$ candidates, 
(middle) $\Xibm\to\jpsi\Xiresm$, $\Xiresm\to\Lz\pi^-_{\rm L}$, and (bottom) $\Xibm\to\jpsi\Xiresm$, $\Xiresm\to\Lz\pi^-_{\rm D}$.
The subscript on the $\pim$ refers to whether the corresponding track is long or downstream.  
The left column shows the combined 7 and 8\tev data and the right one shows the 13\tev data.
The fitted probability distributions functions (PDF) are overlaid.}}
\label{fig:unbinnedFitAll}
\end{figure}

The invariant-mass spectra of selected $\Lb$ and $\Xibm$ candidates are shown in Fig.~\ref{fig:unbinnedFitAll}.
The data are partitioned into the combined 7, 8\tev data samples and the 13\tev data sample, and show the distributions
for $\Lb$ candidates, and $\Xibm$ candidates formed from either long or downstream pions. A simultaneous fit to 
all six distributions is performed in order to determine the signal yields. Each of the signal shapes is described 
by the sum of two Crystal Ball (CB) functions~\cite{Skwarnicki:1986xj} with a common peak position and a common width. 
The tail parameters, which describe the non-Gaussian portion of the signal on either side of the signal peak, are 
independent for the two CB components.  The parameters of the signal shape are determined from large samples of 
simulated signal decays. The background is described by an exponential function, with the shape parameter 
left free in the fit to data. 

The signal-shape fit parameters are: (i) the peak positions, $\bar{m}$, of the $\Lb$ mass in 
the 7,\,8\tev and 13\tev data, (ii) a single mass difference, $\delta m\equiv\bar{m}_{\Xibm}-\bar{m}_{\Lb}$, and
(iii) a scale factor applied to the simulated width of the CB functions, which allows the mass resolution in data to be
slightly different than in simulation.  The values of $\bar{m}_{\Lb}$ are allowed to differ 
for the 7,\,8\tev data and the 13\tev data, since the statistical uncertainty on
each is about four times smaller than the systematic uncertainty from the momentum scale calibration~\cite{LHCb-PAPER-2013-011}.
However, that same calibration renders the corresponding uncertainty on $\delta m$ negligible.

The fitted signal yields and the values of $\bar{m}_{\Lb}$ are shown in Table~\ref{tab:SignalYields}. 
From the fit,  it is determined that
\begin{align*}
\delta m &= 177.30\pm0.39\mevcc, \\
m(\Xibm) &= 5796.70\pm0.39\mevcc,
\end{align*}
\noindent where the uncertainties are statistical only, and we have used $m(\Xibm)=\delta m+m_{\Lb}$, 
with $m_{\Lb}=5619.60\pm0.17$\mevcc~\cite{PDG2018}. The value of $\delta m$ is corrected by $+0.12\pm0.06$\mevcc to
account for a bias observed in the obtained value of $\delta m$, as seen in the fit to large samples of 
simulated signal decays. The uncertainty on this value is due to the size of the simulated samples.

\begin{table*}[tb]
\begin{center}
\caption{\small{Fitted signal yields and peak position of the $\Lb$ signal peak,
as obtained from the fit described in the text. The subscript on the $\pim$ refers to whether the corresponding
track is long or downstream. The uncertainties shown are statistical only.}}
\begin{tabular}{lcc}
\hline\hline
                                             &       $7,\,8$\tev    &      13\tev \\
\hline
\\ [-2.5ex]
$N(\Lb\to\jpsi\Lz)$        &   $~\,13307\pm137$     &    $~\,14793\pm150$ \\
$N(\Xibm\to\jpsi\Xiresm,~\Xiresm\to\Lz\pi^-_{\rm L})$    &    $~~~203\pm16$       &    $~~~258\pm22$     \\
$N(\Xibm\to\jpsi\Xiresm,~\Xiresm\to\Lz\pi^-_{\rm D})$  &    $~~~266\pm20$       &    $~~~357\pm26$     \\
$\bar{m}_{\Lb}$\,(MeV/$\it{c}^{\rm 2}$)                         &  $5619.52\pm0.09$  &    $5619.28\pm0.09$ \\  
\\ [-2.6ex]
\hline\hline
\end{tabular}
\label{tab:SignalYields}
\end{center}
\end{table*}

The ratio of efficiencies in Eq.~\ref{eq:Req} is determined from weighted simulations of the signal decays. The $\Lb$ simulation is weighted in 
bins of ($\eta$, $\pt$) of the $b$ baryon to reproduce the 2D distribution observed in the data, after the background contribution 
is subtracted using the \sPlot method~\cite{Pivk:2004ty}. 
We assume that the $\Xibm$ spectrum is the same as that of the $\Lb$, and variations are investigated when
assessing systematic uncertainties. By studying the distributions of the fraction of the momentum 
carried by the decay products in each part of the decay chain, it is found that the simulation differs
from the corresponding spectra observed in data.
The simulation is weighted to match the distributions observed in data for the momentum ratio $p_{\jpsi}/p_{\Lb}$ 
and the momentum asymmetry $(p_p-p_{\pim})/(p_p+p_{\pim})$ in the $\Lz$ decay. 
After this weighting is applied, a large number of other observables are compared, such as decay times, flight distances, 
$p$, and $\pt$, and good agreement is found between both $\Lb$ and $\Xibm$ data and simulation. For the $\Xibm$
sample, only the $(\eta,\pt)$ and $(p_p-p_{\pim})/(p_p+p_{\pim})$ weights are needed to obtain good agreement with the data.

The resulting efficiencies are summarized in Table~\ref{tab:effs1}. The
efficiencies associated with the detector acceptance, the reconstruction and selection,
and the trigger requirements are given, along with the total selection efficiencies.
The relative efficiency is approximately 14\% for both the 7, 8\tev and 13\tev data sets.
For the 7, 8\tev values, the efficiencies represent the weighted average value.
This small value is due to the combination of the relatively low momentum
and usage of only long tracks for the $\pim$ meson in the $\Xiresm$ decay. 

\begin{table*}[tb]
\begin{center}
\caption{\small{Selection efficiencies as obtained from the simulation of
 $\Lb\to\jpsi\Lz$ and $\Xibm\to\jpsi\Xiresm$, $\Xiresm\to\Lz\pi^-_{\rm L}$ 
decays at $\sqrt{s}=7,\,8$\tev and 13\tev. The efficiencies ($\epsilon$) listed are those
associated with the detector acceptance (acc), the reconstruction and selection (sel), the trigger (trig), 
their product, and the relative efficiency.}}
\begin{tabular}{lcccc}
\hline\hline
                                          &        \multicolumn{2}{c}{7,\,8\tev}             &       \multicolumn{2}{c}{13\tev} \\
Final state                               & $\Lb\to\jpsi\Lz$ & $\Xibm\to\jpsi\Xiresm$ &   $\Lb\to\jpsi\Lz$ & $\Xibm\to\jpsi\Xiresm$ \\
\hline
$\epsilon_{\rm acc}$ (\%)                  &  $18.9\pm0.1$    &  $\,17.3\pm0.1$       &    $19.8\pm0.1$      & $18.0\pm0.1$ \\
$\epsilon_{\rm sel}$ (\%)                  & $~\,2.86\pm0.02$  & $~~0.42\pm0.01$        & $~\,2.91\pm0.01$    & $~\,0.42\pm0.01$ \\
$\epsilon_{\rm trig}$ (\%)                 &  $73.2\pm0.3$     &  $\,75.6\pm0.7$     &   $75.4\pm0.2$  &  $77.8\pm0.5$   \\
\hline
\\ [-2.6ex]
$\epsilon$ ($10^{-2}~\%$)                 &  $39.5\pm0.4$  &  $~~5.56\pm0.11$     &   $43.5\pm0.3$  &  $~\,5.85\pm0.06$   \\
$\epsilon_{\Xibm}/\epsilon_{\Lb}$ (\%)      &    \multicolumn{2}{c}{$14.1\pm0.3$}    &  \multicolumn{2}{c}{$13.4\pm0.2$}  \\
\\ [-2.7ex]
\hline\hline
\end{tabular}
\label{tab:effs1}
\end{center}
\end{table*}

From the signal yields and relative efficiencies, the ratios $R$ are computed to be
\begin{align*}
R &= (10.8\pm0.9)\times10^{-2}~~[\sqrt{s}=7,8\tev], \\
R &= (13.1\pm1.1)\times10^{-2}~~[\sqrt{s}=13\tev],
\end{align*}
\noindent where the uncertainties are statistical only.

The difference between the $\Xibm$ and $\Lb$ baryon production asymmetries is determined using the relation
\begin{align}
A_{\rm prod}(\Xibm)-A_{\rm prod}(\Lb)=\alpha(\Xibm)-\alpha(\Lb)-A_{\rm det}(\pim),
\end{align}
\noindent where $\alpha(\Xibm)$ $[\alpha(\Lb)]$ is the raw yield asymmetry between the
${\Xibm\to\jpsi\Xiresm}$ and ${\Xibbarp\to\jpsi\Xiresbarp}$ $[{\Lb\to\jpsi\Lz}$ and ${\Lbbar\to\jpsi\Lbar}]$ decays.
In the difference of the raw yield asymmetries, the $\Lz$ detection asymmetry cancels since the kinematical 
properties are similar. 
The $\pim$ detection asymmetry, $A_{\rm det}(\pim)$, has been measured~\cite{LHCb-PAPER-2016-013,LHCb-PAPER-2018-010},
and, while it is consistent with zero, an asymmetry of up to about 1\% in this low momentum region cannot be discounted.
In the above expression, it is expected, and assumed, that there is no direct \CP violation in these decays.

The raw yield asymmetries are obtained by fitting for the signal yields separately for 
the beauty baryon and antibaryon subsamples. The fit is similar to that which was described previously,
except that the CB width scale factors are fixed to the values obtained from the fit to the full sample, since the
mass resolution can be assumed to be the same for the $b$ baryons and antibaryons. The fitted signal yields are
shown in Table~\ref{tab:yieldByCharge}, along with the resulting raw asymmetries. 
\begin{table*}[tb]
\begin{center}
\caption{\small{Yields of $\Lb$ and $\Xibm$ decays, split by the charge of the final state, and their asymmetries,
for the combined 7,\,8\tev data samples and the 13\tev data sample. 
Uncertainties are statistical only.}}
\begin{tabular}{lcccc}
\hline\hline
\\ [-2.5ex]
                       &  \multicolumn{2}{c}{$\Lb\to\jpsi\Lz$} &   \multicolumn{2}{c}{$\Xibm\to\jpsi\Xiresm$} \\
                       &       $\Lb$    &        $\Lbbar$      &   $\Xibm$         &  $\Xibbarp$             \\
\hline
$N_{\sqrt{s}=7,\,8\tev}$                  &  $6827\pm94$  &        $6480\pm92$  &   $236\pm18$      &   $230\pm18$             \\
$\alpha_{\sqrt{s}=7,\,8\tev}$     & \multicolumn{2}{c}{$(2.6\pm1.0)\%$} &   \multicolumn{2}{c}{$~~(1.3\pm5.4)\%$}  \\
$N_{\sqrt{s}=13\tev}$      &  $7602\pm102$  &        $7182\pm99$  &   $304\pm21$      &   $326\pm22$         \\
$\alpha_{\sqrt{s}=13\tev}$     & \multicolumn{2}{c}{$(2.8\pm1.0)\%$} &   \multicolumn{2}{c}{$(-3.5\pm4.8)\%$} \\
\hline\hline
\end{tabular}
\label{tab:yieldByCharge}
\end{center}
\end{table*}
The difference in production asymmetries are readily found to be
\begin{align*}
[A_{\rm prod}(\Xibm)-A_{\rm prod}(\Lb)] &= (-1.3\pm5.6)\%~~[\sqrt{s}=7,8\tev], \\
[A_{\rm prod}(\Xibm)-A_{\rm prod}(\Lb)] &= (-6.3\pm4.9)\%~~[\sqrt{s}=13\tev], 
\end{align*}
\noindent where the uncertainties are due to the signal yields obtained in this analysis.

To obtain $A_{\rm prod}(\Xibm)$, previous measurements of $A_{\rm prod}(\Lb)$ at $\sqrt{s}=7$ and 8\tev are used~\cite{LHCb-PAPER-2016-062}.
Since the value of $A_{\rm prod}(\Lb)$ averaged over the LHCb acceptance is not expected to change significantly with center-of-mass energy~\cite{PhysRevD.91.054022},
and the measured values of $A_{\rm prod}(\Lb)$ obtained at $\sqrt{s}=7$ and 8\tev are compatible~\cite{LHCb-PAPER-2016-062}, they are averaged,
taking the systematic uncertainties as fully correlated, to obtain ${A_{\rm prod}(\Lb)=(2.4\pm1.4\pm0.9)\%}$. An alternate measurement of 
$A_{\rm prod}(\Lb)$ yielded results that are consistent with the above value~\cite{LHCb-PAPER-2015-032}.
The value at 7,\,8\tev is also used for the $\Xibm$ asymmetry measurement at 13\tev , and a systematic uncertainty, 
which is discussed below, is assigned. The $\Xibm$ asymmetries are found to be
\begin{align*}
A_{\rm prod}(\Xibm) &= (\phantom{-}1.1\pm5.6)\%~~[\sqrt{s}=7,8\tev],\\
A_{\rm prod}(\Xibm) &= (-3.9\pm4.9)\%~~[\sqrt{s}=13\tev].
\end{align*}

In the mass measurement, most sources of systematic uncertainty cancel, since it relies on the mass difference, $\delta m$.
The modulus of the correction of 0.12\mevcc described previously is assigned as a systematic uncertainty.
The signal shape uncertainty is quantified by performing an alternate fit
using the sum of two Gaussian functions. Apart from a common peak value, all shape parameters are left free in the fit.
The difference with respect to the nominal value, 0.06\mevcc, is assigned as a systematic uncertainty. 
The background shape uncertainty is assessed by using a 
first-order polynomial in place of the nominal exponential function, and is found to change the result 
by 0.01\mevcc. The systematic uncertainties due to the momentum scale and energy loss have been evaluated 
previously~\cite{LHCb-PAPER-2012-048} and are found to contribute 0.01\mevcc each.
Knowledge of the $\Xiresm$ mass contributes an uncertainty of 0.07\mevcc. Adding these uncertainties in quadrature, 
the total systematic uncertainty on $\delta m$ is 0.15\mevcc.

For the measurement of $R$, several sources of uncertainty are considered, which are summarized in Table~\ref{tab:syst}.
The efficiency for all decay products to be within the LHCb acceptance is derived from the simulation, and 
could depend on the polarization of the $\Lb$ or $\Xibm$ baryon.
To investigate this effect, variations in the $\Lb$ and $\Xibm$ polarization are considered, 
including full polarization, zero polarization, and using the helicity amplitudes presented in Ref.~\cite{LHCb-PAPER-2012-057}.
All three variations are found to give statistically compatible acceptance corrections. 
The assigned uncertainty of 3.0\% reflects the
statistical precision of the test.  

The systematic uncertainty due to the signal and background functions is estimated by using
alternate choices for each, as described above for the uncertainty on $\delta m$, leading to
an uncertainty of 2.0\%. The $\Lb$ and $\Xibm$ simulations are weighted as discussed previously and
reproduce well the kinematical distributions of the final-state particles seen in data. 
However, due to low $\Xibm$ signal yields,  
variations with respect to the nominal weighting are considered. In particular, 
a 3\% change in the relative efficiency is seen when applying an additional
weight to the $\Xibm$ pseudorapidity spectrum that is permissible by the data.
A significantly smaller difference is seen when weighting the $\Xibm$ baryon's $\pt$ spectrum.
A 3\% uncertainty is therefore assigned to account for potential differences in the $(\eta,\pt)$ spectrum 
of $\Lb$ and $\Xibm$ baryons.

Uncertainties in the detection efficiency of the $\pim$ meson from the $\Xiresm$ decay
enters directly into the result for the ratio $R$.
The tracking efficiency in simulation has been calibrated using a tag-and-probe method~\cite{LHCb-DP-2013-002} using 
$\jpsi\to\mup\mun$ decays, however, the calibration only covers the kinematic region $p>5$\gevc and $1.9<\eta<4.9$.
Outside this region, no correction to the tracking efficiency in simulation is applied and an uncertainty of 5\% is assigned to the tracking efficiency. 
This value is justified based upon a comparison of
the reconstructed momentum spectrum of $\pim$ mesons from $\Lb\to\jpsi\Lz$ decays in data and simulation, 
where the $\Lz$ baryons are formed from long tracks. These tracks serve as a good proxy for the $\pim$ meson from $\Xiresm$ baryon decay, 
since they also have low momentum and large impact parameter. Averaging over the tracks within and outside
the range covered by the tracking calibration, an uncertainty of 4.5\% on the $\pim$ tracking efficiency is obtained.
As a cross-check, the analysis is repeated using only $\pim$ candidates in the range covered by the calibration, and the 
$R$ values are consistent with the nominal results.

Potential uncertainties due to the $\Xiresm$ mass
requirement may arise from differences in the $\Xiresm$ mass resolution, or possibly a (Cabibbo-suppressed) nonresonant
$\Lz\pim$ contribution. To quantify the potential size of such effects, the $\Xibm$ signal yield in the
$\Xiresm$ sideband region, $10<|M(\Lz\pi^{-}_{\rm L,D})-M(p\pim)+m_{\Lz}-m_{\Xiresm}|<20$\mevcc is measured. The yield in that region, which is 
consistent with zero, is taken as a systematic uncertainty. Other $\Xiresm$ selections are very loose
and are studied by comparing background-subtracted distributions of relevant variables in data with those in simulation.
From the observed differences an uncertainty of 1.4\% is assigned. The uncertainty on $R$ due to the knowledge of the 
$\Xibm$ lifetime, $\tau_{\Xibm}=1.571\pm0.040$\,ps~\cite{PDG2018}, is estimated by weighting the simulation to
replicate 0.04\,ps shorter and longer lifetimes. The effect on $R$ of the $\Lb$ lifetime uncertainty is negligible.
Lastly, the simulated sample sizes contribute 2.0\% uncertainty to the relative efficiency.
\begin{table*}[tb]
\begin{center}
\caption{\small{Summary of relative systematic uncertainties on the production ratio $R$.}}
\begin{tabular}{lcc}
\hline\hline                      
Source                       &        Value (\%)  \\
\hline
\\ [-2.6ex]
$\Lb$, $\Xibm$ polarization  &         3.0     \\
Signal and background shape  &         2.0     \\
$\Xibm$ production spectra   &         3.0 \\ 
$\pim$ tracking efficiency   &         4.5 \\
$\Xiresm$ mass resolution \& non-resonant $\Lz\pim$ &         3.0 \\
$\Xiresm$ selections            &         1.4 \\
$\Xibm$ lifetime             &         0.5 \\
Simulated sample sizes       &         2.0 \\
\hline
Total                        &         7.6 \\
\hline\hline
\end{tabular}
\label{tab:syst}
\end{center}
\end{table*}

The uncertainty on the $\Xibm$ production asymmetry receives contributions from the $\pim$ detection asymmetry and
the measurement of $A_{\rm prod}(\Lb)$. The pion detection asymmetry uncertainty is assigned to be 1\%, as mentioned previously.
Taking the sum in quadrature of the statistical and systematic uncertainties in the value of $A_{\rm prod}(\Lb)=(2.4\pm1.4\pm0.9)\%$,
a 1.7\% systematic uncertainty is assigned. Since the average value of 
$A_{\rm prod}(\Lb)$ at $\sqrt{s}=7$ and 8\tev~\cite{LHCb-PAPER-2016-062} could differ from that at 13\tev~\cite{PhysRevD.91.054022}, an
additional systematic uncertainty of 1.5\% is assigned to the measured value of $A_{\rm prod}(\Xibm)$ at 13\tev.
The total systematic uncertainty in $A_{\rm prod}(\Xibm)$ is therefore 1.9\% and 2.5\% for the 7,\,8\tev and 13\tev data samples, respectively.

In summary, data samples collected at $\sqrt{s}$ = 7, 8 and 13\tev have been used to measure the ratio of production 
rates of $\Xibm$ and $\Lb$ baryons in the pseudorapidity and $\pt$ region, $2<\eta<6$ and $\pt<20$\gevc, to be
\begin{align*}
R &= (10.8\pm0.9\pm0.8)\times10^{-2}~~[\sqrt{s}=7,8\tev], \\
R &= (13.1\pm1.1\pm1.0)\times10^{-2}~~[\sqrt{s}=13\tev],
\end{align*}
\noindent where the uncertainties are statistical and systematic. From the values of $R$,
the ratios of fragmentation fractions are determined to be
\begin{align*}
\frac{f_{\Xibm}}{f_{\Lb}} &= (6.7\pm0.5\pm0.5\pm2.0)\times10^{-2}~~[\sqrt{s}=7,8\tev], \\
\frac{f_{\Xibm}}{f_{\Lb}} &= (8.2\pm0.7\pm0.6\pm2.5)\times10^{-2}~~[\sqrt{s}=13\tev].
\end{align*}
The last uncertainty, due to the assumed SU(3) flavor symmetry and taken to be 30\%, is an estimate of the typical size
of SU(3)-breaking effects between decays related by this symmetry.
The LHCb results show no significant dependence on the center-of-mass energy in the 7 to 13\tev range.
These results are consistent with the predictions in Refs.~\cite{DiWang,Jiang:2018iqa}, which used
production ratio measurements of $\Xibz\to\Xicp\pim$ and $\Lb\to\Lc\pim$ decays at 7 and 8\tev~\cite{LHCb-PAPER-2014-021} 
and an estimated value for $\BR(\Xicp\to p\Km\pip)$. Assuming that $f_{\Xibz}\approx f_{\Xibm}$, these results indicate that in the forward region,
$b$ quarks fragment into $\Xib$ baryons at about 15\% of the rate at which they fragment into $\Lb$ baryons.
Previous measurements of $R$ by the CDF~\cite{Aaltonen:2009ny} and D0~\cite{Abazov:2007am} 
collaborations are about two standard deviations larger than the results reported here, however, those measurements
are performed in $p\bar{p}$ collisions at $\sqrt{s}=2$\tev and in the central rapidity region $|\eta|<2$.

The mass difference, $\delta m$, and the corresponding value of the $\Xibm$ mass, $m(\Xibm)$, are measured to be
\begin{align*}
\delta m &= 177.30\pm0.39\pm0.15\mevcc, \\
m(\Xibm) &= 5796.70\pm0.39\pm0.15\pm0.17\mevcc,
\end{align*}
\noindent where the last uncertainty is due to the $\Lb$ mass. This $\Xibm$ mass measurement 
includes the data used in Ref.~\cite{LHCb-PAPER-2012-048}, and therefore supersedes those results. 
This measurement represents the most precise determination of the $\Xibm$ mass, and is
consistent with the previous most precise measurement of the mass difference of 
$178.36\pm0.46\pm0.16$\mevcc~\cite{LHCb-PAPER-2014-048}. 

The $\Xibm$ production asymmetry is also measured for the first time. The values at the lower and higher
center-of-mass energies, are
\begin{align*}
A_{\rm prod}(\Xibm) &= (\phantom{-}1.1\pm5.6\pm1.9)\%~~[\sqrt{s}=7,8\tev],\\
A_{\rm prod}(\Xibm) &= (-3.9\pm4.9\pm2.5)\%~~[\sqrt{s}=13\tev].
\end{align*}
\noindent 
The asymmetries are consistent with zero at the level of a few percent.



\section*{Acknowledgements}
%
%
\noindent We thank M. Voloshin for interesting and helpful discussions on 
theoretical aspects of this work.
We express our gratitude to our colleagues in the CERN
accelerator departments for the excellent performance of the LHC. We
thank the technical and administrative staff at the LHCb
institutes.
We acknowledge support from CERN and from the national agencies:
CAPES, CNPq, FAPERJ and FINEP (Brazil); 
MOST and NSFC (China); 
CNRS/IN2P3 (France); 
BMBF, DFG and MPG (Germany); 
INFN (Italy); 
NWO (Netherlands); 
MNiSW and NCN (Poland); 
MEN/IFA (Romania); 
MSHE (Russia); 
MinECo (Spain); 
SNSF and SER (Switzerland); 
NASU (Ukraine); 
STFC (United Kingdom); 
NSF (USA).
We acknowledge the computing resources that are provided by CERN, IN2P3
(France), KIT and DESY (Germany), INFN (Italy), SURF (Netherlands),
PIC (Spain), GridPP (United Kingdom), RRCKI and Yandex
LLC (Russia), CSCS (Switzerland), IFIN-HH (Romania), CBPF (Brazil),
PL-GRID (Poland) and OSC (USA).
We are indebted to the communities behind the multiple open-source
software packages on which we depend.
Individual groups or members have received support from
AvH Foundation (Germany);
EPLANET, Marie Sk\l{}odowska-Curie Actions and ERC (European Union);
ANR, Labex P2IO and OCEVU, and R\'{e}gion Auvergne-Rh\^{o}ne-Alpes (France);
Key Research Program of Frontier Sciences of CAS, CAS PIFI, and the Thousand Talents Program (China);
RFBR, RSF and Yandex LLC (Russia);
GVA, XuntaGal and GENCAT (Spain);
the Royal Society
and the Leverhulme Trust (United Kingdom);
Laboratory Directed Research and Development program of LANL (USA).

\clearpage


\addcontentsline{toc}{section}{References}
\bibliographystyle{LHCb}
\bibliography{main,standard,LHCb-PAPER,LHCb-CONF,LHCb-DP,LHCb-TDR}

\newpage
\centerline{\large\bf LHCb collaboration}
\begin{flushleft}
\small
R.~Aaij$^{28}$,
C.~Abell{\'a}n~Beteta$^{46}$,
B.~Adeva$^{43}$,
M.~Adinolfi$^{50}$,
C.A.~Aidala$^{77}$,
Z.~Ajaltouni$^{6}$,
S.~Akar$^{61}$,
P.~Albicocco$^{19}$,
J.~Albrecht$^{11}$,
F.~Alessio$^{44}$
% ...

{\footnotesize \it
% ...
$ ^{6}$Universit{\'e} Clermont Auvergne, CNRS/IN2P3, LPC, Clermont-Ferrand, France\\
% ...
$ ^{11}$Fakult{\"a}t Physik, Technische Universit{\"a}t Dortmund, Dortmund, Germany\\
% ...
$ ^{19}$INFN Laboratori Nazionali di Frascati, Frascati, Italy\\
% ...
$ ^{28}$Nikhef National Institute for Subatomic Physics, Amsterdam, Netherlands\\
% ...
$ ^{43}$Instituto Galego de F{\'\i}sica de Altas Enerx{\'\i}as (IGFAE), Universidade de Santiago de Compostela, Santiago de Compostela, Spain\\
$ ^{44}$European Organization for Nuclear Research (CERN), Geneva, Switzerland\\
% ...
$ ^{46}$Physik-Institut, Universit{\"a}t Z{\"u}rich, Z{\"u}rich, Switzerland\\
% ...
$ ^{50}$H.H. Wills Physics Laboratory, University of Bristol, Bristol, United Kingdom\\
% ...
$ ^{61}$University of Cincinnati, Cincinnati, OH, United States\\
% ...
$ ^{63}$Syracuse University, Syracuse, NY, United States\\
% ...
$ ^{77}$University of Michigan, Ann Arbor, United States, associated to $^{63}$\\
% ...
}
\end{flushleft}

\end{document}